\documentclass[aps,prl,showpacs,twocolumn,floatfix]{revtex4}
\usepackage{amsmath, amsthm, amssymb,graphicx}
\usepackage{subfigure}

\begin{document}

\title{Substrate-induced bandgap in graphene on hexagonal boron nitride}
\author{Gianluca Giovannetti$^{1,2}$, Petr A. Khomyakov$^2$, Geert Brocks$%
^{2}$, Paul J. Kelly$^{2}$ and Jeroen van den Brink$^{1}$}
\affiliation{$^{1}$Instituut-Lorentz for Theoretical Physics,
Universiteit Leiden, P.O.
Box 9506, 2300 RA Leiden, The Netherlands\\
$^{2}$ Faculty of Science and Technology and MESA+ Institute for
Nanotechnology, University of Twente, P.O. Box 217, 7500 AE Enschede,
The Netherlands.}
\date{\today}

\begin{abstract}
We determine the electronic structure of a graphene sheet on top of
a lattice-matched hexagonal boron nitride (h-BN) substrate using
\textit{ab initio} density functional calculations. The most stable
configuration has one carbon atom on top of a boron atom, the other
centered above a BN ring. The resulting inequivalence of the two
carbon sites leads to the opening of a gap of 53 meV at the Dirac
points of graphene and to finite masses for the Dirac fermions.
Alternative orientations of the graphene sheet on the BN substrate
generate similar band gaps and masses. The band gap induced by
the BN surface can greatly improve room temperature pinch-off
characteristics of graphene-based field effect transistors.

\end{abstract}

 \maketitle

\textit{Introduction} Less than 3 years ago it was discovered that graphene
-- a one-atom-thick carbon sheet -- can be deposited on a silicon-oxide
surface by micromechnical cleavage of high quality graphite~\cite%
{Novoselov:sc04}. The graphene flakes are micrometers in size, sufficiently
large to have contacts attached so as to construct field effect transistors
(FETs). Electrical transport measurements made clear that at room
temperature graphene has an electron mobility of at least 10\,000 $\mathrm{%
cm^{2}/Vs}$, a value ten times higher than the mobility of silicon wafers
used in microprocessors~\cite{Novoselov:sc04,Novoselov:nat05,Zhang:nat05}.
The high mobility is not much affected by a field-induced excess of
electrons or holes.

A graphene sheet has a honeycomb structure with two crystallographically
equivalent atoms in its primitive unit cell. Two bands with $p_{z}$
character belonging to different irreducible representations cross precisely
at the Fermi energy at the $K$ and $K^{\prime}$ points in momentum space.
As a result undoped graphene is a zero-gap semiconductor. The linear
dispersion of the bands results in
quasiparticles with zero mass, so-called Dirac fermions. At energies
close to the degeneracy point the electronic states form perfect Dirac
cones. The absence of a gap, preventing the Dirac
fermions from attaining a finite mass and complicating the use of
graphene in electronic devices~\cite{Geim:natm07}, is related to the
equivalence of the two carbon sublattices of graphene.

The relativistic nature of the Dirac fermions gives rise to counterintuitive
phenomena. One, known as the Klein paradox, is that relativistic electrons
exhibit perfect transmission through arbitrarily high and wide potential
barriers. This effect is related to an unwanted characteristic of graphene
FETs, namely that pinch-off is far from complete~\cite{Katsnelson:natp06}.
If one applies a gate voltage so that either holes or electrons are
injected into the graphene sheet, the FET\ is open and its conductivity high.
One can then try to block the current by tuning the gate voltage to move the
graphene layer towards the charge neutrality point where the Fermi energy
coincides with the Dirac points; at this energy the density of states
vanishes and nominally there are no carriers present. However, it turns out
that in spite of the lack of electronic states the conductivity does not
vanish in this case. Rather, it assumes the minimal value $\sigma _{\mathrm{%
min}}=4e^{2}/h$, where $h$ is Planck's constant and $e$ the unit of charge.
Thus even when pinched-off to its maximum the FET still supports an
appreciable electrical current, which is intrinsic to graphene and related
to the fact that the Dirac fermions are massless~\cite%
{Novoselov:nat05,Zhang:nat05,Katsnelson:natp06,Geim:natm07,Heersche:nat07,vandenBrink:natn07}%
.

\textit{Inducing a gap} The poor pinch-off can only be remedied by
generating a mass for the Dirac fermions. A number of possibilities exist to
do so. One is to use bi-layer graphene which will have a gap if the top
and bottom layers are made inequivalent, for instance by applying a bias
potential~\cite{Katsnelson:natp06}. Another is the use of graphene
nanoribbons, where gaps arise from the lateral constriction of the electrons
in the ribbon. The size of the gap then depends on the detailed structure of
the ribbon edges~\cite{Son:prl06,Han:cm07,Chen:cm07}. We investigate an
alternative possibility and consider graphene on a substrate that makes the
two carbon sublattices inequivalent. This breaks the sublattice symmetry
directly, generating an intrinsic and robust mass for the Dirac fermions.

As a substrate, hexagonal boron nitride (h-BN) is a suitable choice
\cite{Suenaga:sc97}. This wide gap insulator has a layered structure very
similar to that of graphene but the two atoms in the unit cell are chemically
inequivalent. Placed on top of h-BN the two carbon sublattices of graphene
become inequivalent as a result of the interaction with the substrate. Our
band structure calculations in the local-density approximation show that a
gap of at least 53 meV -- an energy roughly twice as large as $k_B T$ at
room temperature -- is induced.
This can be compared to graphene on a copper $(111)$ metallic surface where
the gap is found to be much smaller and can even vanish, depending on the
orientation of the graphene sheet.

\begin{figure}
\includegraphics[width=0.82\columnwidth,angle=0]{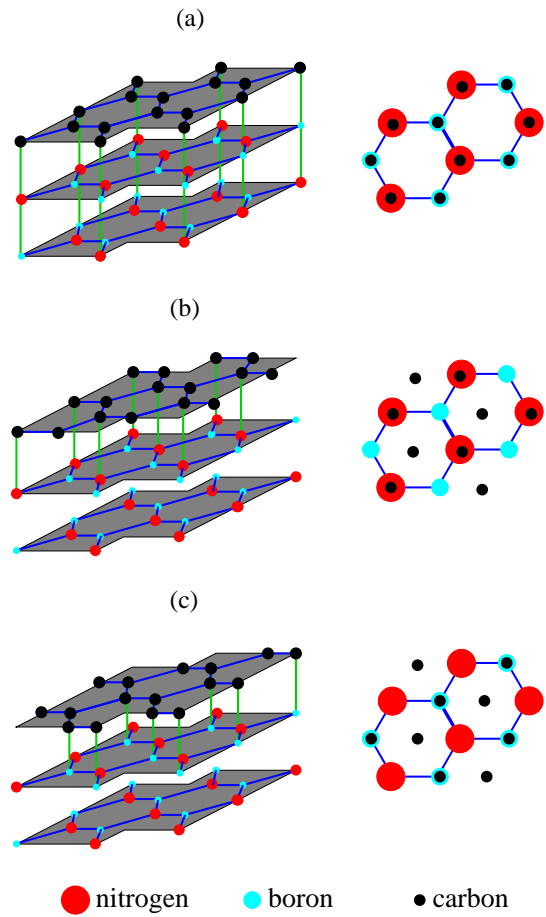}
\caption{(Color online) The three inequivalent orientations of single-layer
graphene on a h-BN surface. Left: sideview, right: topview.}
\label{fig:structures}
\end{figure}

\textit{Stable structure} The lattice mismatch of graphene with
hexagonal boron nitride is less than 2\%. Just as in graphite, the
interaction between adjacent BN layers is weak. The h-BN layers have an
$AA^{\prime }$ stacking: the boron atoms in layer $A$ are directly
above the nitrogen atoms in layer $A^{\prime }$. Within the local
density approximation (LDA) the minimum energy separation of adjacent
layers is found to be 3.24 \AA , which is reasonably close to the
experimental value of 3.33 \AA . Because generalized gradient
approximation (GGA) calculations give essentially no bonding between BN
planes and lead to excessively large values of $c$~\cite{Kern:prb99},
we opt for electronic structure calculations within the LDA.
Electronically, h-BN is a wide gap insulator, with experimentally a gap
of 5.97 eV~\cite{Watanabe:natm04}. This gap is underestimated by about
$33\%$ in LDA. A quasi-particle $GW$ correction on top of the LDA
brings it into very close agreement with experiment
\cite{Blase:prb95,Arnaud:prl06} and reinterprets experiment in terms of
an indirect gap. For the composite graphene layer on top of h-BN
system, we use the LDA lattice parameter for graphene, $a = 2.445$ \AA
.

\begin{figure}
\includegraphics[width=1.035\columnwidth,angle=-0]{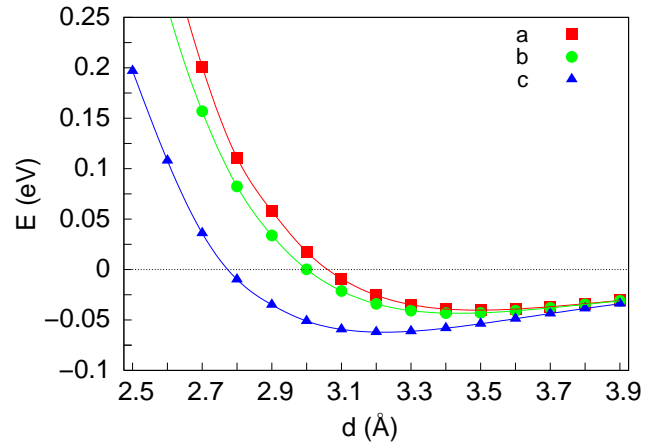}
\caption{(Color online) Total energy $E$ of graphene on h-BN surface
for the three configurations $(a)$, $(b)$, and $(c)$ as a function
of the distance between the graphene sheet and the top h-BN layer.}
\label{fig:energy}
\end{figure}

On the basis of this structural information we construct a unit cell
with 4 layers of h-BN and a graphene top layer. We represent the vacuum
above graphene with an empty space of 12 to 15 \AA . The results to be
presented below converge quickly as a function of the number of h-BN
layers and the width of the vacuum space, consistent with weak interlayer
interactions. No significant difference in the final results were found
when 6 layers of h-BN were used. The in-plane periodicity is that of a
single graphene sheet with a hexagonal unit cell containing two carbon
atoms. We consider three inequivalent orientations of the graphene sheet
with respect to the h-BN, see Fig.~\ref{fig:structures}:\\
- the $(a)$ configuration with one carbon over B, the other carbon over N.\\
- the $(b)$ configuration with one carbon over N, the other carbon
centered above a h-BN hexagon.\\
- the $(c)$ configuration with one carbon over B, the other carbon centered
above a h-BN hexagon.

The self-consistent calculations were performed with the Vienna
Ab-initio Simulation Package (VASP)~\cite{Kresse:prb96,Kresse:cms96}
using a plane wave basis and a kinetic energy cutoff of 600 eV. The
Brillouin Zone (BZ) summations were carried out with the tetrahedron
method and a $36\times 36\times 1$ grid which included the $\Gamma $,
$K$ and $M$ points. A dipole correction avoids interactions between
periodic images of the slab along the $z$-direction
\cite{Neugebauer:prb92}.

The total energies of the three configurations are shown as a function
of the distance between the h-BN surface and the graphene sheet in
Fig.~\ref{fig:energy}. For all distances, the lowest-energy
configuration is $(c)$ with one carbon on top of a boron atom and the
other above a h-BN ring. The equilibrium separation of 3.22 \AA\ for
configuration $(c)$ is smaller than 3.50 \AA\ for configuration $(a)$
and 3.40 \AA\ for configuration $(b)$. For all three configurations the
energy landscape is seen to be very flat around the energy minimum.
Though symmetry does not require inequivalent carbon atoms to be
equidistant from the BN layer, in practice the stiffness of the
graphene sheet prevents any significant buckling.

\begin{figure}
\includegraphics[width=1.11\columnwidth,angle=-0]{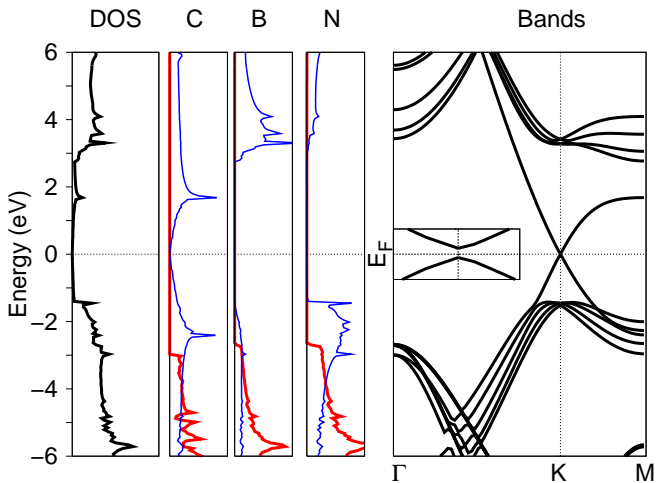}
\caption{(Color online) Band structure along the $\Gamma K$ and $KM$
directions in reciprocal space, total and projected densities of
states (DOS) for the relaxed structure $(c)$ of graphene on h-BN.
Carbon, Boron and Nitrogen projected DOS are shown, with a projection
on the $p$-states in-plane  (red/thick grey lines) and out-of-plane
(blue/thin grey lines). The inset is a
magnification of the bands around the $K$ point, where the gap
opens.} \label{fig:band}
\end{figure}

\textit{Band structure} With the stable structures in hand, we compute
the corresponding electronic band structures and projected densities of
states which are shown in Fig.~\ref{fig:band} for configuration $(c)$.
For the h-BN derived bands a gap of 4.7 eV at the $K$-point is found,
which is nearly identical to the LDA gap value at this particular point
in the Brillouin zone found for bulk h-BN~\cite{Arnaud:prl06}. Within
this boron nitride gap, the bands have entirely carbon character as
expected on the basis of the weak interlayer interactions in both bulk
h-BN and graphite. On the electron-volt scale of Fig.~\ref{fig:band}
the Dirac cone around the $K$-point appears to be preserved. However,
zooming in on that point in the BZ (see inset) reveals that a gap of
53 meV is opened and the dispersion around the Dirac points is
quadratic.

\begin{figure}
\includegraphics[width=0.9\columnwidth,angle=0]{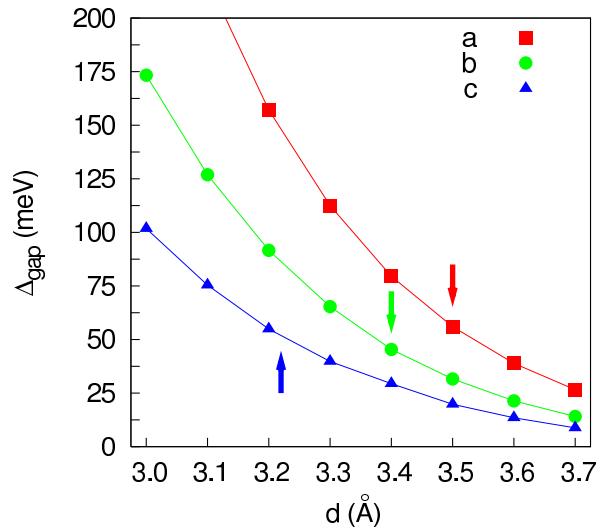}
\caption{(Color online) The values of the gaps for the three
configurations $(a)$, $(b)$, and $(c)$ as a function of the distance
between the graphene sheet and the top h-BN layer. The calculated
equilibrium separations are indicated by vertical arrows.}
\label{fig:gaps}
\end{figure}

The band gaps for the three different configurations are shown in
Fig.~\ref{fig:gaps} as a function of the distance between the graphene
sheet and the h-BN surface. Decreasing this distance increases the gap,
as expected for a physical picture based upon a symmetry-breaking
substrate potential.
The band gaps that are opened at the equilibrium geometries of the $(a)$
and $(b)$ configurations are 56 meV and 46 meV, respectively, which are
comparable to the band gap obtained for configuration $(c)$. The largest
gap is found for the $(a)$ configuration with one carbon atom above a
boron atom and the other above a nitrogen atom. Again, this is expected
for gap opening induced by breaking the symmetry of the two carbon atoms.
Since LDA generally underestimates the gap, the values that we obtain put 
a  lower bound on the induced band gaps, which we thus find to be significantly 
larger than $k_B T$ at room temperature.

Although the lattice mismatch between graphene and h-BN constants is
less than 2\% and can be neglected in a first approximation, in a real
system incommensurability will occur and we expect the strong in-plane
bonding of both graphene and h-BN to prevail over the weak inter-plane
bonding. For graphene on Ir(111) where the lattice mismatch is
$\sim 10\%$, Moir\'{e} patterns have been observed in STM images
\cite{NDiaye:prl06}. There, first-principles calculations showed that
regions could be identified where the graphene was in registry with the
underlying substrate in high symmetry configurations analogous to the
$(a)$, $(b)$ and $(c)$ configurations discussed above, and transition
regions with little or no symmetry \cite{NDiaye:prl06}. The graphene
separation from the substrate varied across the surface leading to
bending of the graphene sheet. If we could take the lattice mismatch
into account in a large supercell in a similar fashion, some areas of
graphene would be forced into the higher energy $(a)$ and $(b)$
configurations with larger separations to the BN substrate. However,
the corresponding band gaps are all of the order of the 50 meV we find
for the lowest energy $(c)$ configuration, or higher. It seems
reasonable to conclude that the broadening resulting from lattice
mismatch will not reduce the gap substantially. 

\textit{Cu(111) substrate} The situation changes markedly for graphene
on a Cu(111) surface. The copper surface layer forms a triangular
lattice, matching that of graphene to better than 4\%. We consider
two configurations of
graphene on Cu(111). Either the center of each carbon hexagon is on
top of a Cu atom, which we call the symmetric configuration in the
following, or every second carbon atom is on top of a Cu atom, which
we call the asymmetric configuration. For the asymmetric and symmetric
configurations, LDA calculations yield equilibrium separations of $3.3$
\AA\ and $3.4$ \AA\ which are comparable to those of graphene on h-BN
(Fig.~\ref{fig:energy}). The total energy difference between the two
configurations is only about $9$ meV. In the asymmetric configuration
a small gap of $11$ meV is opened in the graphene band structure,
whereas in the symmetric configuration the gap remains very close to
zero. In both cases we find very little mixing between copper and
carbon states. The difference between the gaps can be explained
by the fact that the symmetric configuration preserves the graphene
symmetry in the top Cu surface layer, whereas the symmetry is broken
in the asymmetric configuration. The effect of this symmetry breaking
is small, however, and the resulting band gap is much smaller than that
induced by h-BN and comparable to the typical thermal broadening
reported in experiments \cite{Novoselov:nat05,Geim:natm07}. Taking into
account the graphene-Cu lattice mismatch in for instance a super cell 
calculation \cite{NDiaye:prl06} will not to change this conclusion.

For both configurations of graphene on Cu, a charge rearrangement at the
interface is found which moves the Fermi level away from the induced
gap \cite{metalGpaper:2007} by much more than the magnitude of the
gap itself. This is in contrast to a h-BN substrate, where 
the Fermi level remains in the induced gap. Around the Fermi level of 
graphene on Cu the band dispersion is still linear. Consequently, the properties 
characteristic of graphene which result from the linear dispersion should be 
preserved. In for instance tunneling
experiments that require adsorption of graphene on a metallic (Cu) substrate
\cite{Oshima:jpcm97} one should still be able to observe the intrinsic linear 
electronic structure of graphene near the Fermi energy, but no longer at the 
Dirac points. 

\textit{Conclusions} Our density functional calculations show that the
carbon atoms of a graphene sheet preferentially orient themselves directly 
above the boron atoms of a h-BN substrate, with one carbon sublattice 
above the boron sublattice and the other carbon centered above a h-BN ring.
Although graphene interacts only weakly with the h-BN substrate, even
when a few angstroms away the presence of h-BN induces a band gap of 
53 meV, generating an effective mass for the Dirac fermions of  
$4.7 \cdot 10^{-3} \ m_e$, where $m_e$ is the electron mass.
The gap that opens at the Dirac points is
considerably larger than the one for graphene on Cu(111). Additional
quasi-particle interactions, for instance taken into account within a
$GW$ scheme, will increase the value of the gap. The opening of a band gap in
graphene on h-BN offers the potential to improve the characteristics
of graphene-based FETs, decreasing the minimum conductance by orders
of magnitude. Other interesting features such as the valley degree of
freedom, which is related to the degeneracy of the $K$ and $K^{\prime }$
points in the Brillouin zone, remain intact and can still be used to
control an electronic device~\cite{Rycerz:natp07}.
Also the half-integer quantum Hall effect -- a peculiar characteristic of
graphene -- remains unchanged~\cite{Novoselov:nat05,Peres06,Gusynin:prl05}.

\acknowledgments

This work was financially supported by ``NanoNed'', a nanotechnology
programme of the Dutch Ministry of Economic Affairs and by the
``Nederlandse Organisatie voor Wetenschappelijk Onderzoek (NWO)''
via the research programs of ``Chemische Wetenschappen (CW)'' and
the ``Stichting voor Fundamenteel Onderzoek der Materie (FOM)''.
Part of the calculations were performed with a grant of computer
time from the ``Stichting Nationale Computerfaciliteiten (NCF)''.


\begin{thebibliography}{22}
\expandafter\ifx\csname natexlab\endcsname\relax\def\natexlab#1{#1}\fi
\expandafter\ifx\csname bibnamefont\endcsname\relax
  \def\bibnamefont#1{#1}\fi
\expandafter\ifx\csname bibfnamefont\endcsname\relax
  \def\bibfnamefont#1{#1}\fi
\expandafter\ifx\csname citenamefont\endcsname\relax
  \def\citenamefont#1{#1}\fi
\expandafter\ifx\csname url\endcsname\relax
  \def\url#1{\texttt{#1}}\fi
\expandafter\ifx\csname urlprefix\endcsname\relax\def\urlprefix{URL }\fi
\providecommand{\bibinfo}[2]{#2}
\providecommand{\eprint}[2][]{\url{#2}}

\bibitem[{\citenamefont{Novoselov et~al.}(2004)\citenamefont{Novoselov, Geim,
  Morozov, Jiang, Zhang, Dubonos, Grigorieva, and Firsov}}]{Novoselov:sc04}
\bibinfo{author}{\bibfnamefont{K.~S.} \bibnamefont{Novoselov}},
  \bibinfo{author}{\bibfnamefont{A.~K.} \bibnamefont{Geim}},
  \bibinfo{author}{\bibfnamefont{S.~V.} \bibnamefont{Morozov}},
  \bibinfo{author}{\bibfnamefont{D.}~\bibnamefont{Jiang}},
  \bibinfo{author}{\bibfnamefont{Y.}~\bibnamefont{Zhang}},
  \bibinfo{author}{\bibfnamefont{S.~V.} \bibnamefont{Dubonos}},
  \bibinfo{author}{\bibfnamefont{I.~V.} \bibnamefont{Grigorieva}},
  \bibnamefont{and} \bibinfo{author}{\bibfnamefont{A.~A.}
  \bibnamefont{Firsov}}, \bibinfo{journal}{Science}
  \textbf{\bibinfo{volume}{306}}, \bibinfo{pages}{666} (\bibinfo{year}{2004}).

\bibitem[{\citenamefont{Novoselov et~al.}(2005)\citenamefont{Novoselov, Geim,
  Morozov, Jiang, Katsnelson, Grigorieva, Dubonos, and
  Firsov}}]{Novoselov:nat05}
\bibinfo{author}{\bibfnamefont{K.~S.} \bibnamefont{Novoselov}},
  \bibinfo{author}{\bibfnamefont{A.~K.} \bibnamefont{Geim}},
  \bibinfo{author}{\bibfnamefont{S.~V.} \bibnamefont{Morozov}},
  \bibinfo{author}{\bibfnamefont{D.}~\bibnamefont{Jiang}},
  \bibinfo{author}{\bibfnamefont{M.~I.} \bibnamefont{Katsnelson}},
  \bibinfo{author}{\bibfnamefont{I.~V.} \bibnamefont{Grigorieva}},
  \bibinfo{author}{\bibfnamefont{S.~V.} \bibnamefont{Dubonos}},
  \bibnamefont{and} \bibinfo{author}{\bibfnamefont{A.~A.}
  \bibnamefont{Firsov}}, \bibinfo{journal}{Nature}
  \textbf{\bibinfo{volume}{438}}, \bibinfo{pages}{197} (\bibinfo{year}{2005}).

\bibitem[{\citenamefont{Zhang et~al.}(2005)\citenamefont{Zhang, Tan, Stormer,
  and Kim}}]{Zhang:nat05}
\bibinfo{author}{\bibfnamefont{Y.~B.} \bibnamefont{Zhang}},
  \bibinfo{author}{\bibfnamefont{Y.~W.} \bibnamefont{Tan}},
  \bibinfo{author}{\bibfnamefont{H.~L.} \bibnamefont{Stormer}},
  \bibnamefont{and} \bibinfo{author}{\bibfnamefont{P.}~\bibnamefont{Kim}},
  \bibinfo{journal}{Nature} \textbf{\bibinfo{volume}{438}},
  \bibinfo{pages}{201} (\bibinfo{year}{2005}).

\bibitem[{\citenamefont{Geim and Novoselov}(2007)}]{Geim:natm07}
\bibinfo{author}{\bibfnamefont{A.~K.} \bibnamefont{Geim}} \bibnamefont{and}
  \bibinfo{author}{\bibfnamefont{K.~S.} \bibnamefont{Novoselov}},
  \bibinfo{journal}{Nature Materials} \textbf{\bibinfo{volume}{6}},
  \bibinfo{pages}{183} (\bibinfo{year}{2007}).

\bibitem[{\citenamefont{Katsnelson et~al.}(2006)\citenamefont{Katsnelson,
  Novoselov, and Geim}}]{Katsnelson:natp06}
\bibinfo{author}{\bibfnamefont{M.~I.} \bibnamefont{Katsnelson}},
  \bibinfo{author}{\bibfnamefont{K.~S.} \bibnamefont{Novoselov}},
  \bibnamefont{and} \bibinfo{author}{\bibfnamefont{A.~K.} \bibnamefont{Geim}},
  \bibinfo{journal}{Nature Physics} \textbf{\bibinfo{volume}{2}},
  \bibinfo{pages}{620} (\bibinfo{year}{2006}).

\bibitem[{\citenamefont{Heersche et~al.}(2007)\citenamefont{Heersche,
  Jarillo-Herrero, Oostinga, Vandersypen, and Morpurgo}}]{Heersche:nat07}
\bibinfo{author}{\bibfnamefont{H.~B.} \bibnamefont{Heersche}},
  \bibinfo{author}{\bibfnamefont{P.}~\bibnamefont{Jarillo-Herrero}},
  \bibinfo{author}{\bibfnamefont{J.~B.} \bibnamefont{Oostinga}},
  \bibinfo{author}{\bibfnamefont{L.~M.~K.} \bibnamefont{Vandersypen}},
  \bibnamefont{and} \bibinfo{author}{\bibfnamefont{A.~F.}
  \bibnamefont{Morpurgo}}, \bibinfo{journal}{Nature}
  \textbf{\bibinfo{volume}{446}}, \bibinfo{pages}{56–} (\bibinfo{year}{2007}).

\bibitem[{\citenamefont{van~den Brink}(2007)}]{vandenBrink:natn07}
\bibinfo{author}{\bibfnamefont{J.}~\bibnamefont{van~den Brink}},
  \bibinfo{journal}{Nature Nanotechnology}, \textbf{\bibinfo{volume}{2}},
  \bibinfo{pages}{199–} (\bibinfo{year}{2007}).

\bibitem[{\citenamefont{Son et~al.}(2006)\citenamefont{Son, Cohen, and
  Louie}}]{Son:prl06}
\bibinfo{author}{\bibfnamefont{Y.~W.} \bibnamefont{Son}},
  \bibinfo{author}{\bibfnamefont{M.~L.} \bibnamefont{Cohen}}, \bibnamefont{and}
  \bibinfo{author}{\bibfnamefont{S.~G.} \bibnamefont{Louie}},
  \bibinfo{journal}{Phys. Rev. Lett.} \textbf{\bibinfo{volume}{97}},
  \bibinfo{pages}{216803} (\bibinfo{year}{2006}).

\bibitem[{\citenamefont{Han et~al.}(2007)\citenamefont{Han, Ozyilmaz, Zhang,
  and Kim}}]{Han:cm07}
\bibinfo{author}{\bibfnamefont{M.~Y.} \bibnamefont{Han}},
  \bibinfo{author}{\bibfnamefont{B.}~\bibnamefont{Ozyilmaz}},
  \bibinfo{author}{\bibfnamefont{Y.}~\bibnamefont{Zhang}}, \bibnamefont{and}
  \bibinfo{author}{\bibfnamefont{P.}~\bibnamefont{Kim}},
  \bibinfo{note}{cond-mat/0702511}.

\bibitem[{\citenamefont{Z.~Chen et~al.}(2007)\citenamefont{Z.~Chen, Rooks, and
  Avouris}}]{Chen:cm07}
\bibinfo{author}{\bibfnamefont{Y.~M.~L.} \bibnamefont{Z.~Chen}},
  \bibinfo{author}{\bibfnamefont{R.~J.} \bibnamefont{Rooks}}, \bibnamefont{and}
  \bibinfo{author}{\bibfnamefont{P.}~\bibnamefont{Avouris}},
  \bibinfo{note}{cond-mat/0701599}.

\bibitem[{\citenamefont{Suenaga et~al.}(1997)\citenamefont{Suenaga, Colliex,
  Demoncy, Loiseau, Pascard, and Willaime}}]{Suenaga:sc97}
\bibinfo{author}{\bibfnamefont{K.}~\bibnamefont{Suenaga}},
  \bibinfo{author}{\bibfnamefont{C.}~\bibnamefont{Colliex}},
  \bibinfo{author}{\bibfnamefont{N.}~\bibnamefont{Demoncy}},
  \bibinfo{author}{\bibfnamefont{A.}~\bibnamefont{Loiseau}},
  \bibinfo{author}{\bibfnamefont{H.}~\bibnamefont{Pascard}}, \bibnamefont{and}
  \bibinfo{author}{\bibfnamefont{F.}~\bibnamefont{Willaime}},
  \bibinfo{journal}{Science} \textbf{\bibinfo{volume}{278}},
  \bibinfo{pages}{653} (\bibinfo{year}{1997}).

\bibitem[{\citenamefont{Kern et~al.}(1999)\citenamefont{Kern, Kresse, and
  Hafner}}]{Kern:prb99}
\bibinfo{author}{\bibfnamefont{G.}~\bibnamefont{Kern}},
  \bibinfo{author}{\bibfnamefont{G.}~\bibnamefont{Kresse}}, \bibnamefont{and}
  \bibinfo{author}{\bibfnamefont{J.}~\bibnamefont{Hafner}},
  \bibinfo{journal}{Phys. Rev. B} \textbf{\bibinfo{volume}{59}},
  \bibinfo{pages}{8551} (\bibinfo{year}{1999}).

\bibitem[{\citenamefont{Watanabe et~al.}(2004)\citenamefont{Watanabe,
  Taniguchi, and Kanda}}]{Watanabe:natm04}
\bibinfo{author}{\bibfnamefont{K.}~\bibnamefont{Watanabe}},
  \bibinfo{author}{\bibfnamefont{T.}~\bibnamefont{Taniguchi}},
  \bibnamefont{and} \bibinfo{author}{\bibfnamefont{H.}~\bibnamefont{Kanda}},
  \bibinfo{journal}{Nature Materials} \textbf{\bibinfo{volume}{3}},
  \bibinfo{pages}{404} (\bibinfo{year}{2004}).

\bibitem[{\citenamefont{Blase et~al.}(1995)\citenamefont{Blase, Rubio, Louie,
  and Cohen}}]{Blase:prb95}
\bibinfo{author}{\bibfnamefont{X.}~\bibnamefont{Blase}},
  \bibinfo{author}{\bibfnamefont{A.}~\bibnamefont{Rubio}},
  \bibinfo{author}{\bibfnamefont{S.~G.} \bibnamefont{Louie}}, \bibnamefont{and}
  \bibinfo{author}{\bibfnamefont{M.~L.} \bibnamefont{Cohen}},
  \bibinfo{journal}{Phys. Rev. B} \textbf{\bibinfo{volume}{51}},
  \bibinfo{pages}{6868} (\bibinfo{year}{1995}).

\bibitem[{\citenamefont{Arnaud et~al.}(2006)\citenamefont{Arnaud, Lebegue,
  Rabiller, and Alouani}}]{Arnaud:prl06}
\bibinfo{author}{\bibfnamefont{B.}~\bibnamefont{Arnaud}},
  \bibinfo{author}{\bibfnamefont{S.}~\bibnamefont{Lebegue}},
  \bibinfo{author}{\bibfnamefont{P.}~\bibnamefont{Rabiller}}, \bibnamefont{and}
  \bibinfo{author}{\bibfnamefont{M.}~\bibnamefont{Alouani}},
  \bibinfo{journal}{Phys. Rev. Lett.} \textbf{\bibinfo{volume}{96}},
  \bibinfo{pages}{026402} (\bibinfo{year}{2006}).

\bibitem[{\citenamefont{Kresse and
  Furthmuller}(1996{\natexlab{a}})}]{Kresse:prb96}
\bibinfo{author}{\bibfnamefont{G.}~\bibnamefont{Kresse}} \bibnamefont{and}
  \bibinfo{author}{\bibfnamefont{J.}~\bibnamefont{Furthmuller}},
  \bibinfo{journal}{Phys. Rev. B} \textbf{\bibinfo{volume}{54}},
  \bibinfo{pages}{11169} (\bibinfo{year}{1996}{\natexlab{a}}).

\bibitem[{\citenamefont{Kresse and
  Furthmuller}(1996{\natexlab{b}})}]{Kresse:cms96}
\bibinfo{author}{\bibfnamefont{G.}~\bibnamefont{Kresse}} \bibnamefont{and}
  \bibinfo{author}{\bibfnamefont{J.}~\bibnamefont{Furthmuller}},
  \bibinfo{journal}{Comp. Mat. Sci.} \textbf{\bibinfo{volume}{6}},
  \bibinfo{pages}{15} (\bibinfo{year}{1996}{\natexlab{b}}).

\bibitem[{\citenamefont{Neugebauer and Scheffler}(1992)}]{Neugebauer:prb92}
\bibinfo{author}{\bibfnamefont{J.}~\bibnamefont{Neugebauer}} \bibnamefont{and}
  \bibinfo{author}{\bibfnamefont{M.}~\bibnamefont{Scheffler}},
  \bibinfo{journal}{Phys. Rev. B} \textbf{\bibinfo{volume}{46}},
  \bibinfo{pages}{16067} (\bibinfo{year}{1992}).

\bibitem[{\citenamefont{{N'Diaye} et~al.}(2006)\citenamefont{{N'Diaye},
  Bleikamp, Feibelman, and Michely}}]{NDiaye:prl06}
\bibinfo{author}{\bibfnamefont{A.~T.} \bibnamefont{{N'Diaye}}},
  \bibinfo{author}{\bibfnamefont{S.}~\bibnamefont{Bleikamp}},
  \bibinfo{author}{\bibfnamefont{P.~J.} \bibnamefont{Feibelman}},
  \bibnamefont{and} \bibinfo{author}{\bibfnamefont{T.}~\bibnamefont{Michely}},
  \bibinfo{journal}{Phys. Rev. Lett.} \textbf{\bibinfo{volume}{97}},
  \bibinfo{pages}{215501} (\bibinfo{year}{2006}).

\bibitem[{\citenamefont{Giovannetti et Al}(2007)}]{metalGpaper:2007}
\bibinfo{author}{\bibfnamefont{G.}~\bibnamefont{Giovannetti}},
\bibinfo{author}{\bibfnamefont{P. A.}~\bibnamefont{Khomyakov}},
\bibinfo{author}{\bibfnamefont{G.}~\bibnamefont{Brocks}},
\bibinfo{author}{\bibfnamefont{J.}~\bibnamefont{van den Brink}},
\bibinfo{author}{\bibfnamefont{P. J.}~\bibnamefont{Kelly}},
\bibinfo{journal}{to be published}.

\bibitem[{\citenamefont{Oshima and Nagashima}(1997)}]{Oshima:jpcm97}
\bibinfo{author}{\bibfnamefont{C.}~\bibnamefont{Oshima}} \bibnamefont{and}
  \bibinfo{author}{\bibfnamefont{A.}~\bibnamefont{Nagashima}},
  \bibinfo{journal}{J. Phys.: Condens. Matter.} \textbf{\bibinfo{volume}{9}},
  \bibinfo{pages}{1} (\bibinfo{year}{1997}).

\bibitem[{\citenamefont{Rycerz et~al.}(2007)\citenamefont{Rycerz, Tworzydlo,
  and Beenakker}}]{Rycerz:natp07}
\bibinfo{author}{\bibfnamefont{A.}~\bibnamefont{Rycerz}},
  \bibinfo{author}{\bibfnamefont{J.}~\bibnamefont{Tworzydlo}},
  \bibnamefont{and} \bibinfo{author}{\bibfnamefont{C.~W.~J.}
  \bibnamefont{Beenakker}}, \bibinfo{journal}{Nature Physics}
  \textbf{\bibinfo{volume}{3}}, \bibinfo{pages}{172} (\bibinfo{year}{2007}).

\bibitem{Peres06}
N.M.R. Peres, F. Guinea and A. H. Castro Neto, Phys. Rev. B {\bf 73}, 125411 (2006). 

\bibitem[{\citenamefont{Gusynin and Sharapov}(2005)}]{Gusynin:prl05}
\bibinfo{author}{\bibfnamefont{V.~P.} \bibnamefont{Gusynin}} \bibnamefont{and}
  \bibinfo{author}{\bibfnamefont{S.~G.} \bibnamefont{Sharapov}},
  \bibinfo{journal}{Phys. Rev. Lett.} \textbf{\bibinfo{volume}{95}},
  \bibinfo{pages}{146801} (\bibinfo{year}{2005}).

\end{thebibliography}

\end{document}